# GRAPESPH: Cosmological SPH simulations with the special purpose hardware GRAPE

Matthias Steinmetz

Max–Planck–Institut für Astrophysik, Karl–Schwarzschild–Straße 1, 85740 Garching b. München, FRG

## Abstract

A combined N–body/hydrodynamical code is presented. Hydrodynamical properties are determined using smoothed particle hydrodynamics (SPH). The gravitational interaction of gas and collisionless particles is treated in a direct summation approach which benefits from the high speed of the special purpose hardware GRAPE (GRAvity PipE). Besides gravitational forces, GRAPE also returns the list of neighbours and can, therefore, be used to speed up the hydrodynamical part, too. After the interaction list has been passed, density, pressure forces, propagation and interpolation of particles etc. are calculated on the front end, a 50 MHz SUN SPARC 10. In order to combine SPH and GRAPE, possible limitations due to the hardware design of GRAPE are carefully analyzed and modifications compared to current SPH codes are discussed.

The resulting code, GRAPESPH is similarly flexible as TREESPH. 50-55% of the CPU time is spent to calculate the densities and the pressure forces on the front end, 15-20% to calculate gravity, and about 10% in miscellaneous subroutines. Another 20% are required by communication, mainly to read out the neighbour list via the VME interface. The main shortcoming is the inflexible, hardwired force law (a Plummer law), which makes it difficult to include periodic boundary conditions. Also because of the limited dynamic range, GRAPESPH seems to be less suitable to perform large scale structure simulations, where the resolution should be high everywhere in the simulation volume. The resulting code seems, however, especially well suited to investigate the formation of individual objects in a large scale structure environment, like e.g., galaxies or clusters. Such simulations require a very high spatial resolution, but only within a relatively small subvolume. By means of a multiple time step scheme, time step constraints due to local stability criteria can almost be avoided. The total performance is at least half as good as TREESPH on a CRAY and for most applications it seems to be even better. The CPU time per time step is only slightly dependent on the clustering state. GRAPESPH, therefore, provides a very attractive alternative to the use of supercomputers in cosmology.

## 1. Introduction

During the last two decades N–body simulations have shed much light on the formation of large scale structure, galaxy clusters and galaxies. Besides analytical methods, like the Press–Schechter formalism (Press & Schechter 1974, Bond et al. 1991, Lacey & Cole 1993), they have become the major tool for theoretical investigations in extragalactic astronomy. These simulations, however, only follow the evolution of the distribution of dark matter, which is commonly believed to dominate the dynamics on large scales ($\gtrsim 10\,\mathrm{Mpc}$), but which, by definition, cannot be directly observed. During the last few years one has begun to extend the pure N–body simulations to include also gas dynamical effects. Pioneering work was done by Evrard (1988), Hernquist & Katz (1989), Navarro & Benz (1991), Cen (1992) and Umemura (1993). These advanced simulations allowed for a more realistic description of the dynamics on small scales ($< 5\,\mathrm{Mpc}$), which is affected or even dominated by gas dynamical effects. Moreover, on larger scales the combined gas dynamical and N–body simulations provide the link between the dark matter and directly observable quantities.

Most cosmological hydrodynamical simulations have been performed using *smoothed particle hydrodynamics* (SPH, Lucy 1977). Indeed, SPH seems to become *the* hydrodynamical method for extragalactic investigations, mainly because it is (i) intrinsically three dimensional, (ii) a Lagrangian method and (iii) able to provide surprisingly accurate results in some applications even with a small number of particles. Furthermore, the SPH Euler equation is similar to that of the gravitational N–body system, additionally modified by a short range force term. Form this point of view, SPH seems to be the natural extension of the N-body approach. Recently, SPH became even more attractive, because it is very well suited for GRAPE (Sugimoto et al. 1990), a special purpose hardware, which efficiently solves the gravitational N–body problem, and which can be connected to standard workstations via a VME interface. GRAPE allows one to perform simulations on workstations, which before required a supercomputer like e.g., a CRAY (Steinmetz 1994). Note, however, that the largest formal resolution ($\gtrsim 270^3$ particles) has been used in the codes of Cen & Ostriker (1992), Ryu et al. (1993) and Bryan et al. (1995), which are based on a finite difference scheme. Furthermore, finite difference methods using a Lagrangian description (Gnedin 1995), as well as moving mesh schemes (Pen 1995) and adaptive mesh refinement (Berger & Oliger 1984, Berger & Colella 1989) seem to provide attractive alternatives to SPH. For a more detailed analysis of the capabilities and limits of SPH compared to finite difference schemes we refer to Steinmetz & Müller (1993, henceforth SM93) for general hydrodynamical aspects and to Kang et al. (1994) for cosmological applications.

In this paper we describe a SPH code designed to study the formation of galaxies and galaxy clusters. The code, originally based on a binary tree code (SM93), is rewritten to run efficiently in combination with the special purpose hardware GRAPE. The outline of the paper is the following: In a first section we describe the basics of SPH and some modification like kernel steepening and an artificial viscosity which (almost) vanishes in pure shear flows. Section 3 resumes the main properties of the special purpose hardware GRAPE, and analyzes the hardware design in order to evaluate its suitability for SPH. Section 4 continues with a detailed description of the adaption of SPH to run on GRAPE. In Section 5, the performance of such a code is investigated in detail and compared with other SPH codes running on traditional computers. However, we do not repeat a detailed description of the outcome of different test problems, because the transition from TREESPH to GRAPESPH changes only little on the capabilities and limits of SPH as thoroughly discussed in Evrard (1988), Hernquist & Katz (1989), SM93, Navarro & White (1993) or Kang et al. (1994). The paper concludes with a summary and a discussion on well suited applications for GRAPESPH.

## 2. Basics of Smoothed Particle Hydrodynamics

In smoothed particle hydrodynamics (SPH, Lucy 1977), each physical quantity $A(\mathbf{r})$ is approximated by a smoothed estimate $\langle A(\mathbf{r}) \rangle$ given by

$$\langle A(\mathbf{r}) \rangle = \int d^3 r' A(\mathbf{r}') W(\mathbf{r} - \mathbf{r}', h) \approx \sum_i m \frac{A(\mathbf{r}_i)}{\varrho(\mathbf{r}_i)} W(\mathbf{r} - \mathbf{r}_i, h) \,, \tag{1}$$

where $W$ is the smoothing kernel, i.e., a function strongly peaked at $|\mathbf{r} - \mathbf{r}'| = 0$ similar to a Gaussian (see below). The so-called smoothing length $h$ and the particle number determine the numerical resolution of a simulation. Approximating the integral in Eq. (1) by a summation can be interpreted as a Monte–Carlo integration of $\langle A(\mathbf{r}) \rangle$ (Lucy 1977). Most frequently the kernel $W$ is given by the so-called $B_2$-spline (Monaghan 1985)

$$W(\mathbf{r} - \mathbf{r}', h) = \frac{8}{\pi h^3} \begin{cases} 1 - 6 \cdot u^2 + 6 \cdot u^3 \,, & 0 \leq u \leq \frac{1}{2} \\ 2 \cdot (1 - u)^3 \,, & \frac{1}{2} < u \leq 1 \\ 0 \,, & u > 1 \end{cases} \tag{2}$$

with $u = |\mathbf{r} - \mathbf{r}'|/h$. For the later discussion of the neighbour search, it is convenient to compactify the kernel on the interval $[0;1]$, instead of on the interval $[0;2]$ (Monaghan 1985, and most other authors working on SPH). The main advantages of the spline kernel are, that (i) it gives a second–order accurate interpolation, (ii) it is positive definite, and (iii) it has compact support. For later use we stress that there is no principal argument against the use of different kernels or softening lengths for different quantities.

The disadvantage of the spline kernel (and similarly also of Gaussian type kernels) is its vanishing gradient at $u = 0$, i.e., the pressure gradient decreases with increasing compression. In numerical simulations carried out by Schüssler & Schmidt (1981), who assumed an isothermal, not self–gravitating gas, an artificial formation of dense knots of gas was observed in the case of a kernel with vanishing gradient. This behaviour can be avoided by means of a cusp–like kernel with a finite gradient at $u = 0$. More recently, Swegle, Hicks & Attaway (1995) demonstrated by a von Neumann stability analysis, that this clumping is due to an instability of the kernel against compression, if $d^2W/dx^2$ is negative. However, the vanishing gradient at $u = 0$ is essential to provide a smooth interpolation, and hence is a desired property of $W$. In the test problems investigated by SM93 the clumping effect was not observed, because due to the adiabaticity of the considered flows the pressure gradient rises with increasing compression. Therefore, particle distances less than $\frac{1}{3}h$, where the second derivative of the $B_2$ spline becomes negative, are rarely encountered.

To overcome the clumping instability Thomas & Couchman (1992) proposed to leave the kernel unchanged, but to modify its gradient for $u < \frac{1}{3}$ according to

$$\frac{dW}{du}(u) = \frac{dW}{du}(u = \frac{1}{3}) = -\frac{16}{\pi h^3} , u \leq \frac{1}{3} . \tag{3}$$

Although this form is only marginally stable for $u \leq \frac{1}{3}$, due to particles with distances in the range $\frac{1}{3} < u \leq 1$ a pressure gradient is provided which grows with increasing compression. Our test simulations with the modified kernel show indeed a much more stable and less clumpy behaviour. Moreover, as a pleasant side effect, a larger time step can be used, which saves considerable amounts of computing time. 1D shock tube simulations also exhibit a much sharper shock representation and a reduction of post shock oscillations especially for small particle numbers.

There are several, formally equivalent ways to derive the hydrodynamical equations for the smoothed estimates $\langle A(\mathbf{r}) \rangle$ from Eq. (1) (for a review, see e.g., Monaghan 1985, 1992). According to a comprehensive set of test calculations (SM93), the best results are obtained for the following formulation:

$$\begin{aligned}
\frac{d\mathbf{v}_i}{dt} &= -\sum_j m \left( \frac{P_i}{\varrho_i^2} + \frac{P_j}{\varrho_j^2} + Q_{ij} \right) \nabla_i W(r_{ij}, h) \\
&\quad - G \sum_j \frac{m_j}{r_{ij}^3} 4\pi \int_0^{r_{ij}} dr\, r^2\, W(r, h)(\mathbf{r}_i - \mathbf{r}_j)
\end{aligned} \tag{4}$$

$$\frac{d\varepsilon_i}{dt} = \frac{P_i}{\varrho_i^2} \sum_j m (\mathbf{v}_i - \mathbf{v}_j) \cdot \nabla_i W(r_{ij}, h) + \frac{1}{2} \sum_j m_j Q_{ij} (\mathbf{v}_i - \mathbf{v}_j) \cdot \nabla_i W(r_{ij}, h), \tag{5}$$

where $r_{ij} = |\mathbf{r}_i - \mathbf{r}_j|$. Note that due to the antisymmetry of the force between two particles $i$ and $j$, linear and angular momentum are conserved. Note, that the densities in Eq. (4-5) involve another summation (Eq. 1, with $A$ substituted by $\varrho$). Since the force on particle $i$ also depends on the density of $j$, both equations cannot be solved within one loop. The list of neighbours, which contribute to the sum in Eq. (1,4-5), has therefore either to be stored for every particle, or to be calculated twice.

The Monaghan–Gingold tensor $Q_{ij}$ mimics an artificial viscosity and is introduced to handle shocks avoiding post-shock oscillations (Monaghan & Gingold 1983). It is given by

$$Q_{ij} = \begin{cases} \frac{-\alpha c_{ij}\mu_{ij} + \beta \mu_{ij}^2}{\varrho_{ij}}, & (\mathbf{r}_i - \mathbf{r}_j) \cdot (\mathbf{v}_i - \mathbf{v}_j) \leq 0 \\ 0, & \text{otherwise,} \end{cases} \quad (6)$$

$$\mu_{ij} = \frac{h_{ij}(\mathbf{v}_i - \mathbf{v}_j) \cdot (\mathbf{r}_i - \mathbf{r}_j)}{(r_{ij}^2 + \eta^2)}, \quad (7)$$

where $c_{ij}$, $h_{ij}$ and $\varrho_{ij}$ are the arithmetic means of the sound velocity $c$, the smoothing length $h$, and the density $\varrho$, respectively. The parameter $\eta \approx 0.1h$ prevents the Monaghan–Gingold tensor $Q_{ij}$ from becoming infinite, if $|\mathbf{v}_i - \mathbf{v}_j| \neq 0$ and $r_{ij} = |\mathbf{r}_i - \mathbf{r}_j|$ tends to zero. The parameters $\alpha$ and $\beta$ mimic a first and second Navier-Stokes viscosity coefficient. Usually we set $\alpha = 0.5$ and $\beta = 1$, but for problems involving strong shocks the choice $\alpha = 1$ and $\beta = 2$ is more appropriate to avoid post-shock oscillations. An increased artificial viscosity, however, increases the smearing of shock waves, and also reduces the Courant time step (Monaghan & Gingold 1983).

As noted by Hernquist & Katz (1989), the artificial viscosity given by Monaghan–Gingold tensor does not vanish in pure shear flows. Katz & Gunn (1991) argued that this property may disturb a forming disk, and used a formulation of the artificial viscosity, which is directly based on the SPH–estimate of $\nabla \cdot \mathbf{v}$. Although the amount of shear viscosity can be reduced with this artificial viscosity, our experience shows (see SM93), however, that it smears out strong shock waves much more than the Gingold–Monaghan viscosity. Furthermore, in our simulations (Steinmetz & Müller 1995) we did not observe a severe disk perturbation due to the shear viscosity for particle numbers in the disk of about 4000. However, decreasing the particle number below 1000, angular momentum transport due to the shear viscosity seems to become a severe problem, i.e., the typical transport time scale becomes shorter than the Hubble time.

These problems can be almost completely prevented by means of a modified artificial viscosity tensor $\widetilde{Q}_{ij}$ similar to that proposed by Balsara (1995):

$$\widetilde{Q}_{ij} = Q_{ij}(f_i + f_j)$$

$$f_i = \frac{|\langle \nabla \cdot \mathbf{v} \rangle_i|}{|\langle \nabla \cdot \mathbf{v} \rangle_i| + |\langle \nabla \times \mathbf{v} \rangle_i| + 0.0001 c_i/h_i}, \quad (8)$$

where the term $0.0001 c_i/h_i$ is introduced to prevent divergences. In case of a shear-free, compressive flow, i.e., $\nabla \cdot \mathbf{v} \neq 0$, $\nabla \times \mathbf{v} = 0$ and, therefore, $f = 1$, the viscosity is identical to that proposed by Monaghan & Gingold (1983). However, in the presence of shear flows, the viscosity is diminished. The suppression factor $f$ is an order of magnitude estimate of the irrotational fraction of the flow. In case of a pure shear flow, i.e., $\nabla \cdot \mathbf{v} = 0$, $\nabla \times \mathbf{v} \neq 0$, $f$ vanishes, the viscosity is completely suppressed. Using this formulation, the unphysical angular momentum transport could be reduced significantly. In case of the original Monaghan & Gingold viscosity the half angular momentum radius of a galactic size gaseous disk, which consists of 280 particles, grew by a factor of 2 within 3 Gyr. After the modification, the half angular momentum radius varied by less than 10% within a Hubble time. This observation also agrees with that of Navarro (private communication).

Similar to other SPH codes (Hernquist & Katz 1989, Benz 1990), the time integration is done with a multiple time step scheme based on a binary hierarchy of time bins. The smoothing length $h$ is allowed to vary in space and time as described in SM93. Within this paper particles in the hierarchy of time bins for which the force has to be calculated will be called *active particles*, whereas positions of *passive particles*, which only have to be updated on higher time levels, are obtained via interpolation.

## 3. GRAPE hardware specifications

GRAPE (=GRAvity PipE) is a special purpose computer designed to calculate efficiently the gravitational N–body problem in a direct summation approach (Sugimoto et al. 1990). The currently distributed version, GRAPE 3A-f, consists of a board of roughly 27x37 cm (6U Eurocard) containing 8 N–body integrator chips. The chips calculate the forces exerted by up to 131072 particles with positions $\mathbf{x}_j$ at 8 arbitrary positions $\mathbf{x}_i$ in parallel. The force law is hardwired to

$$\frac{d\mathbf{v}_i}{dt} = -\sum_{j=1}^{N} \frac{m_j}{(|\mathbf{r}_i - \mathbf{r}_j|^2 + \varepsilon_i^2)^{1.5}} (\mathbf{r}_i - \mathbf{r}_j) \,, \qquad (9)$$

i.e., a Plummer law. The sustained performance of a board is 4.8 Gflop/s. The board is connected via a VME interface to a standard workstation e.g., a SUN SPARC 10. A C and FORTRAN library provides the software interface between GRAPE and the frontend. In addition to the direct summation approach, meanwhile a treecode (Makino & Funato 1993) and, more recently, a P3M code (Brieu, Summers & Ostriker 1995) were developed for GRAPE. Beside the force and potential, GRAPE also returns an index array containing all neighbours within a sphere $h_i$ around $\mathbf{x}_i$. This feature makes GRAPE especially attractive to combine it with SPH (Umemura et al. 1993).

To clarify the following discussion, we will use the index $i$ for a particle for which the force is calculated, and the index $j$ for those particles $j$ which exert the force. $N$ is the total number of particles, $N_n^i$ the number of neighbours within a sphere of radius $h_i$. The array $k^i(l), l = 1, N_n^i$ provides the indices of the neighbours.

In order to achieve a high speed, concessions had to be made concerning the accuracy of the force calculation. For collisionless N–body systems, the force accuracy has to be not more accurately calculated than up to a few percent, which roughly corresponds to the accuracy of a tree code. Therefore, instead of IEEE 4 or 8 byte real numbers, GRAPE internally works with a 20 bit fixed point number format for positions, a 56 bit fixed point number format for forces, and a 14 bit logarithmic number format for masses (Okumura et al. 1993). The rescaling from 4 or 8 byte real to the internal number format and back is done by the library. The user has to supply only the minimum mass and length scale. Besides the limited force accuracy one has to keep in mind the following additional hardware limitations:

1. The force law is hardwired to be a Plummer law. It is, therefore, not straightforward to modify the code to handle periodic boundary conditions, since the Ewald corrected force (Hernquist, Bouchet & Suto 1991) as well as the PP force in P3M (Brieu, Summers & Ostriker 1995) are different from a Plummer law.

2. The dynamic range is limited to $\mathtt{minx}\cdot[-2^{18};2^{18}]$ and the mass range to $\mathtt{minm}\cdot[1;64\cdot\varepsilon/\mathtt{minx}]$. Here, $\mathtt{minx}$ and $\mathtt{minm}$ are the minimum resolved length and mass scale, respectively.

3. The board provides all neighbours within a sphere of radius $h_i$ around the particle $i$, but not all those particles $j$ which overlap with their softening $h_j$ the particle $i$. According to the nomenclature of Hernquist & Katz (1989) all neighbours required for a gather formulation are provided, but it is not guaranteed that all neighbours required for the scatter formulation or a combination of both can be found. Therefore, if the smoothing length of the particle $i$ is larger, but that of the particle $j$ smaller than the interparticle distance, the particle $i$ contributes to the force on particle $j$ but not vice versa. Newton's third law is violated resulting in a non conservation of linear and angular momentum.

4. The neighbour hardware recognizes a particle $j$ to be a neighbour of $i$, if $|\mathbf{r}_i - \mathbf{r}_j|^2 + \max(\varepsilon_i, \mathtt{minx})^2 < \max(h_i, \mathtt{minx})^2$. Furthermore, one also has to take into account the relatively large round off error of a few percent.

5. The FIFO chip which stores the neighbour list is limited to 1023 neighbours for all 8 chips together. However, only disjunct neighbours count. If a particle $j$ is neighbour to several of the particles $i$ which are loaded on the 8 chips, it occupies only one memory location. GRAPE returns only the list of disjunct neighbours and a mask which specifies which chips share a neighbour. The actual neighbour list for every chip is than deconvolved on the front end (Makino, private communication). To give an example: if two chips have the neighbour list (8,11,15,47) and (8,15,42), GRAPE returns the disjunct list (8,11,15,42,47) and the mask (binary) (11,10,11,01,10). With other words, in the worst case, every particle $i$ has a completely different neighbour list and *all 8 particles i together* cannot have more than 1023 neighbours. If, however, the neighbour lists of the 8 particles $i$ are identical, *every particle* can have up to 1023 neighbours. On average, the possible length of the interaction list can, therefore, be increased, if particles $i$ are grouped in such a way that they have similar interaction lists. Furthermore, the neighbour hardware tends to give incorrect results, if all of the 8 chips have no neighbour at all. In this pathological case, for every chip a neighbour list is returned which consists of one particle with index 17.

6. The neighbour search starts at $N$ and counts down to 1. If the neighbour list exceeds 1023, all further potential neighbours are ignored. In the index array $k^i(l), l = 1\ldots N_n^i, i = 1,8)$ returned from GRAPE, the index array is ordered $k^i(1) < k^i(2)\ldots < k^i(N_n^i)$ for every particle $i$. Due to these properties, one knows that in the case of an overflow of the neighbour list all neighbours with indices larger than $\min_{i=1,8}(k^i(1))-1$ are found.

In spite of the restrictions imposed by the hardware, a combination of GRAPE and SPH nevertheless seems to be intriguing: For usual SPH codes, which are based on tree or P3M methods, the SPH force and density calculation requires only 5-20% of the CPU time, most of the computing time being used up in order to calculate the gravitational force and the interaction list. Assuming that GRAPE can perform the latter two operations much faster, a maximum speedup between a factor of 5 to 20 can be expected compared to the front end. Therefore, in combination with a SPARC 10, timings can be expected which up to now were only reached by supercomputers. Because the high speed of GRAPE makes the use of a treecode or P3M unnecessary, the memory requirement can also be reduced by a factor larger than two. Since SPH simulations usually are CPU time limited rather than memory limited, the typically small (in comparison to supercomputers) memory of workstations should not pose a severe problem. Furthermore, in the case of a multiple time step scheme, the smallest time bins usually have a low occupation number. Although the overhead related to tree construction etc. is small if the forces are calculated for all particles, its relative importance is much larger for the sparsely occupied low time levels. Since this overhead potentially vanishes in the GRAPE approach, the multiple time step scheme becomes even more efficient.

## 4. GRAPSPH, a SPH code for GRAPE

### 4.1 Dynamic range and boundary conditions

Generally, the limited dynamic range and the restricted force law, which make it difficult to implement periodic boundary conditions, disfavour GRAPESPH for large scale structure simulations, where the same numerical resolution is required everywhere in the simulation volume. GRAPESPH is more competitive in the domain of TREESPH than in that of P3MSPH. It is very well suited to investigate the formation of single galaxies or groups and clusters (e.g., Katz & Gunn 1991, Thomas & Couchman 1992, Katz & White 1993, Navarro & White 1993, 1994, Navarro, Frenk & White 1995, Steinmetz & Müller 1994, 1995), but seems to be less suitable for simulations of the gas distribution on large scales (e.g., Cen et al. 1990, Cen & Ostriker 1992, Katz, Weinberg & Hernquist 1992). In order to include the tidal field and mass inflow due to large scale modes, we apply the multi–mass–technique as e.g. described in Katz & White (1993), Navarro & White (1994) or Bartelmann, Steinmetz & Weiss (1995): Based on a coarsely grained, large scale N–body simulation, single haloes are picked out. Every halo is then re-calculated in a high–resolution run using the same large-scale density modes, but adding smaller–scale modes up to the Nyquist frequency of the refined grid. The mass distribution outside the refined region is approximated by a few thousand macro–particles, whose mass is increasing with distance.

These massive boundary particles, however, cause conflicts with the restricted range of length scales and masses: Typically, nearby boundary particles have masses similar to particles in the high resolution region. The masses of the most distant particles, however, can be more massive by a factor of a few thousands. In order to avoid an over-/underflow of interparticle forces, the masses must not vary by more than a factor of about 100. We solve this problem by grouping particles of similar mass (e.g., within one decade) together and calculate the forces exerted by every group separately, the corresponding mass and length scales being appropriately set.

### 4.2 Force antisymmetry

To ensure energy and momentum conservation, the gravitational softening is fixed in time and space, but different for dark matter, gas, stars and boundary particles. To guarantee force antisymmetry (at least for the gravitational forces), the interaction of particles of different groups is computed with their arithmetically averaged softening $(\varepsilon_i + \varepsilon_j)/2$.

Up to now we have not found an easy solution to ensure antisymmetry of the pressure forces. Of course, the solution is evident, if one omits the multiple time step scheme: every pair of particle $ij$ with $\Delta r < \max(h_i, h_j)$ is found either when the force on $i$ or on $j$ is calculated. This cannot be guaranteed in a multiple time step scheme, because, e.g., the particle $j$ might not belong to the active particles. However, the gain in performance obtained in the multiple time step scheme can exceed a factor of ten for strongly cooling systems like e.g., galaxies (see below), a speed up which one does not want to sacrifice.

On the other hand, treecodes violate conservation laws, too, mainly for the same reason. Furthermore, the antisymmetry of the interparticle forces only holds, because the $\nabla h$ terms are neglected. A more rigorous derivation of the equations of motion including the $\nabla h$ terms arising from a variable smoothing length also produces non antisymmetric interparticle forces (Villumsen, private communication). Hence, we tried the following approach, which cannot guarantee antisymmetric forces but seems to provide meaningful results: for every particle $i$ all neighbours within a sphere of $1.3\,h_i$ are considered to be potential neighbours, i.e., all neighbours with $h_j < 1.3\,h_i$ are included. Whenever $\varrho(\mathbf{r})$ varies only smoothly in $\mathbf{r}$, $h(\mathbf{r})$ is smooth, too. Only in case of very strong density and $h$ gradients some neighbours can be missed. However, since the gradient of the kernel scales with $h^{-4}$ these particles only contribute little to the force on $i$ and their neglect causes an error which is probably smaller than e.g., the error arising from the neglect of the $\nabla h$ terms. The main disadvantage of this technique is that on average 50% of the potential neighbours (for highly clustered states even more) can be thrown away (i.e., particles with $\Delta r_{ij} < 1.3\,h_i$ and $\Delta r_{ij} > \max(h_i, h_j)$). However, before being discarded, for these particles one has to calculate the interparticle distance, too and has to compared it with $\max(h_i, h_j)$. Hence, they also contribute to the CPU time on the front end and to the communication time.

## 4.3 Adaptive Smoothing and Number of Neighbours

The probably most severe problem for GRAPESPH is the limited number of neighbours. On the first glance, the situation seems to be not too bad: In the case of a homogeneous lattice and a smoothing length equal to the grid spacing, every particle has 33 neighbours, 6 of them having weight zero. However the 30% tolerance mentioned above to find mutual pairs of neighbours doubles the number of potential neighbours. Additionally, it was recently shown by a stability analysis (Morris 1994), that a usually assumed combination of a spline kernel and a smoothing length equal to half the grid spacing (recall the different definition of the kernel in Eq. 2) is dynamically unstable. This instability can be avoided by either using a Gaussian kernel, a quartic or quintic spline, or a factor of 1.2 larger smoothing. In all of these cases, the number of neighbours is increased by a factor 1.5 to 2.5, i.e., dangerously close to the limit of 128 particles per chip on average.

Last, but not least, the gas in simulations of galaxies and galaxy clusters is strongly affected by cooling processes. With the currently included physics and the insufficient numerical resolution, the collapse only stops artificially due to the limited gravitational softening. Therefore, several hundreds of particles can accumulate within the gravitational softening $\varepsilon$.

The number of neighbours is also connected to the scheme with which the smoothing length is adapted in space and time. In Hernquist & Katz (1989), the number of neighbours was fixed to $\approx 40$ allowing only a small tolerance of a few percent. Besides the fact that 40 neighbours seem to be too small to provide dynamical stability, in our opinion, such a restrictive scheme can also give rise to pathological situations: Consider a dense knot of gas consisting of 39 particles in an environment of diffuse gas. Just by chance, an environment particle might approach the dense knot of gas. Then, the smoothing lengths of the gas particles in the gas knot will change according to the motion of only one particle, and the density of the gas knot scales with $(\Delta r)^{-3}$, $\Delta r$ being the distance of the environment particle to the gas knot. Physically, this scaling seems to be not very plausible.

In the approach of Benz (1990), Evrard (1988) or SM93, the smoothing is coupled to the gas density or a smoothed estimate of it, which almost excludes pathological situations as above. The disadvantage is that the number of neighbours can vary a lot, a minimum or maximum neighbour number cannot be guaranteed. We modified the scheme described in SM93 in the following way: As long as the number of neighbours is between 50 and 80 the original scheme is used (i.e., $\frac{dh}{dt} \propto \nabla \mathbf{v}$). If the number of neighbours $N_n$ lies between 30 and 50 or between 80 and 120, an additional term $\frac{dh_c}{dt}$ is switched on which tries to reduce the number of neighbours back to the interval [50; 80]:

$$\frac{dh_c}{dt} = \frac{1}{\tau} \begin{cases} (r_{50} - h_{uc}), & N_n < 50 \\ (h_{uc} - r_{80}), & N_n > 80 \\ 0 & 50 \leq N_n \leq 80 \end{cases} \quad (10)$$

where $r_{50}$ and $r_{80}$ are the distances to the 50th and 80th closest neighbour, respectively. $h_{uc}$ is smoothing length according to the original scheme, $\tau$ the time scale on which the gas density is changing. If the number of neighbours exceeds 120 (or falls below 30), $h$ is set equal to the radius of the 50th (80th) closest neighbour. This very restrictive step is only seldom necessary, because usually the number of neighbours can be kept in the range [30;120] without problems, in most cases $N_n$ is in the range [45;85].

Finally, it is still an open question, how to treat smoothing if the mean interparticle distance drops below $\varepsilon$. One way is to restrict the smoothing length to the gravitational softening length, another way is to let it float freely. From the theoretical point of view there is no strong argument in favour of or against both approaches (see above). From the computational point of view, the free floating approach is advantageous: The Courant–Friedrich–Levy time step decreases $\propto h$, but the number of neighbours decreases $\propto h^3$. Considering the limited size of the neighbour list, the free float approach is obviously better suited for GRAPE.

In simulations including feedback processes due to star formation we observed, that in the case of the free floating approach, $h(\mathbf{r})$ saturates at about $\frac{\varepsilon}{5} \ldots \frac{\varepsilon}{10}$. In pure gas dynamical simulations this saturation does not occur and $h$ collapses until it gets into conflict with the lower resolution limit of GRAPE. We, therefore, force the smoothing length to be larger than $h_{\min} = \varepsilon/5 \ldots \varepsilon/10$. If the minimum smoothing is reached, the number of neighbours can become much larger than 120. This situation, however only occurs in the case of very dense and tightly bound gas clumps. Since the size of these high density regions is determined by the gravitational softening, the high formal resolution (in terms of particles per gas knot) is meaningless. Therefore, particles with $h(\mathbf{r}) = h_{\min}$ and more than $\approx 200$ neighbours are combined to new, more massive particles. This reduction process is performed whenever the whole system is synchronized on the largest time level. The particles are grouped by means of a friends–to–friends group finding algorithm with a linking length of $\frac{1}{10} h_{\min}$, i.e., only very tightly bound gas knots are reduced. To avoid relaxation effects, the mass of the combined particle may not exceed the mass of a dark matter particle. We found that by means of these modifications, the average number of neighbour never exceeds 80. These modifications led to a reduction of the CPU time of a simulation by 10–20%.

### 4.4 Neighbour Search

A possible way to handle the neighbour list and its overflows is shown in the code fragment in Appendix A. To compactify the listing, we assume that all SPH–particles are analyzed, and that the number of particles nsph is a multiple of the number of chips nchip. Furthermore, we omit type declaration, dimension statements, bound checking etc. Note, that arrays on GRAPE start at index 0!

First GRAPE is initialized. The scaling factors for mass (minm) and positions (minx) are specified as well as the gravitational softening and the number of gas particles nsph. Afterwards, particles and masses are transmitted to the board. If nsph is larger than 131072, the SPH particles must be successively processed in chunks of 131072 particles. Specifying the gravitational softening to zero (or more precisely to minx) indicates that the force call is only used to find the neighbours. The gravitational force and potential are separately calculated.

Next, the by 30% enlarged smoothing lengths of nchip particles are loaded and the neighbour list is reset to zero. The integer istart will be used later, if the neighbour hardware overflows. Then, the GRAPE force calculation begins, which will deliver the neighbour list. In the first iteration all gas particles are considered (i.e., the IF statement is true), and every particle should find at least one neighbour, namely the particle itself. If, however, an overflow occurs, a second iteration is necessary, where only a subset of particles is considered (nsub < nsph; ELSE). Now it is possible, that none of the chips will detect a neighbour, a case which, as mentioned above, is incorrectly handled by the hardware. Therefore, we add the particle of chip 1 to the subset particles, i.e., at least one neighbour is found and the hardware works correctly. This particle is discarded once the neighbour list is read out.

Next, the neighbour list has to be read out and possible overflows have to be handled. If the number of disjunct neighbours ntot does not exceed 1023, the neighbour list can be read out immediately and added to the list from previous iterations. Density, forces etc. can then be calculated.

If the neighbour list overflows, the (incomplete) neighbour list is read out and the particle nsub+1, which was added to enforce at least one neighbour, is discarded. According to the construction of the neighbour hardware, all neighbours with indices higher than kmin = min(k(istart(i)),i=1,nchip)), i.e., within the interval [kmin; nsph], have been found. Only the subset of particles with indices [1;kmin-1] has still to be considered. These particles are processed in the next iteration. Neighbours found in that iteration will be appended to the neighbour list, i.e., istart has to be increased by nt, the number of neighbours found within the current iteration. Note, that because the neighbours are successively found starting with the highest indices and counting downwards, one has to reset only the particle number, but one has not to send particle positions and masses, i.e., there is practically no communication time involved.

Finally, we have to explain, why active particles are ordered in their $x$–position, before the neighbour search starts. We recall point (5) in the last section, namely that GRAPE delivers the list of disjunct neighbours and the mask to deconvolve that list. Hence, for every disjunct neighbour 4 words are read out, 1 word for the disjunct neighbour list, 1 word for the mask, and both items once more for cross checking the neighbour list. If we assume an average number of neighbours of about 120 (including the 30% tolerance mentioned above), for every gas particle 480 words or 3.75 kByte have to be transmitted. In the case of a 30 000 gas particle simulation and an effective transfer rate of 3 MByte/s via the VME interface, a communication time of about 40 sec results, which is a non–negligible fraction of the total CPU time per time step ($\approx 150$ sec). Furthermore, in SPH one needs the neighbour list twice, for the calculation of the density and of the pressure force, which cannot be done in parallel (see above). Hence, if one wants to avoid to store the neighbour lists of all particles, one has to perform the neighbour search and the communication twice. However, if one could arrange the particles in such a way that particles with neighbouring indices have similar interaction lists, one could reduce the communication time dramatically. This rearrangement must not be CPU intensive. A convenient, efficient and inexpensive way is to sort the particles according to their x coordinate. This procedure is also invoked in some group finding algorithms. By this simple technique, we were able to reduce the communication time for two read outs of the neighbour list to less than 20% of the CPU time per time step. For comparison: In the optimum case, when all particles have absolutely identical interaction lists, the communication time would amount $\approx 10\%$. Only for the first few time steps, when all particles are close to their original position, a regular lattice, a slight performance degradation can be observed.

|       | 65100 particles |       |       |       | 11500 particles |       |       |       |
|-------|---------|-------|-------|-------|---------|-------|-------|-------|
| Model | density | hydro | grav. | misc. | density | hydro | grav. | misc. |
| 1 | 35.3% | 36.4% | 16.5% | 11.8% | 38.1% | 37.4% | 15.6% | 8.9% |
| 2 | 31.7% | 33.5% | 26.9% | 7.9%  | 38.3% | 39.3% | 17.7% | 4.7% |
| 3 | 36.5% | 35.2% | 16.1% | 12.2% | 41.3% | 39.2% | 12.4% | 7.2% |

Table 1: Percentage of CPU times spent in different subroutines for two simulation with different particle numbers. The datasets 1–3 correspond to different subset for the analysis: 1) The whole simulation, 2) for redshift interval [20; 5] when nearly all particles have the same time step, 3) for a fiducial redshift interval [0.26-0.19] when time steps vary by a factor of 500.

## 5. Performance Analysis

### 5.1 The simulations

In order to get realistic estimates for the performance and the load distribution for the different subroutines, we have performed two simulations of the hierarchical formation of galaxies similar to that described in Navarro & White (1994). Two simulations have been performed. One uses 32000 gas, 32000 dark matter, and 1100 boundary particles, while the second one involved 5200 gas, 5200 dark matter, and 1100 boundary particles. The simulations start at a redshift of 21, and the background cosmogony is a $b = 1.7$, $H_0 = 50 \,\mathrm{km\,s^{-1}\,Mpc^{-1}}$, $\Omega = 1$ CDM scenario. A baryon fraction of 5% is assumed. The masses of a gas and collisionless particles in the high resolution run is $4.9 \; 10^6 \,\mathrm{M_\odot}$ and 9.3 $10^7$ M$_\odot$, respectively. The corresponding masses of the low resolution run are by a factor of 6 larger. Gravitational softening was taken to be 1.25 kpc and 2.5 kpc (2.5 kpc and 5 kpc) in the high (low) resolution run. A detailed description of the simulations will be presented elsewhere (Navarro & Steinmetz, in preparation).

For comparison with other cosmological SPH codes, we use timings for simulations of the collapse of a top hat sphere assuming vacuum boundaries (Katz 1992, Steinmetz & Müller 1994, 1995) which partially were done with our TREESPH–code (SM93) on a CRAY, and partially with GRAPESPH. The reference simulation included 4 000 gas particles, 4 000 dark matter particles, and up to 25 000 star particles.

## 5.2 Timings for a complete time step

Timings are analyzed a) for the whole simulation, b) very early in the simulation, when all particles share the same time level and c) late in the simulation, when a highly clustered state develops and the particles are distributed over up to 8 time levels. We distinguish between subroutines required to calculate the SPH-density, the pressure force, the gravitational force and miscellaneous subroutines. Timings were obtained with an implementation of GRAPESPH, where the interaction list is not stored but calculated twice. The timings are summarized in Tab. 1 and in Fig. 1 and 2. One can see that on average about 70-75% of the CPU time is used to almost equal parts for the determination of the densities and the pressure forces. About 15-20% are required for the gravity, and about 10% are used up by miscellaneous subroutines.

**Dependence on the environment** As an intriguing result we found, that the CPU time of GRAPESPH only slightly depends on the environment and the clustering state. Whereas a complete force calculation required 140 sec for a homogeneous environment, it typically took 190 sec in the highly clustered state at the end of the simulation. This is a much weaker dependence than in case of TREESPH or P3MSPH where timings commonly vary by a factor of five or even more. Even the timings of adaptive mesh schemes (Couchman, Thomas & Pearce 1995) vary by a factor of about 3. Of course, the attractive scaling of GRAPESPH is an immediate consequence of the direct summation approach, where the gravitational force evaluation is independent of the clustering state. The time variation only arises from an on average larger length of the interaction list. For a homogeneous environment, on average 120 particles can be found in the search radius of $1.3h_i$, and 60 particles contribute to the density and to the pressure force. In the highly clustered system near the end of the simulation, on average 70 particles contribute to SPH quantities, but the number of neighbours within $1.3h_i$ has increased to about 200. The almost a factor of 2 larger communication time and the fact that for false neighbours, at least $|r_i - r_j|$ has to be calculated, before they can be discarded, result in a 35% increase of the CPU time.

**Dependence on particle number** Theoretically one would expect, that the CPU time per time step increases with particle number between $N$ and $N^2$ depending on whether the short range SPH-forces ($\propto N$) or the long ranged gravitational forces ($\propto N^2$) dominate. We found that the higher resolved simulation took about a factor of 8.5 longer per time step than the low resolution run. It, therefore, scales slightly steeper than linear (expected factor 6). The subroutines in which density and pressure forces are calculated slowed down by a factor 7.5 to 8 (expected 6), that of the gravity by a factor of 15 (expected 36). All these scalings are almost independent of the clustering state. For 5000 particles, the communication time ($\propto N$) is no longer negligible compared to the time for the force calculation. Therefore, the $N^2$ scaling is not yet reached, which explains the lower than expected slow down of the gravity subroutine. The stronger slow down of the density and pressure force calculation can partially be explained, because the force call ($\propto N^2$) required to get the neighbour list becomes more important. Furthermore, we found empirically that in the higher resolved simulation on average more particles can be found within the search radius $1.3\,h_i$, i.e., the higher resolved simulation is probably able to achieve a more pronounced clustering state. Finally, current Risc architectures suffer from performance degradations if the data cannot be kept within the processor cache, which is more likely for larger particle numbers.

The CPU time required to carry out a complete simulation scales with the particle number $\propto N^{4/3}$. The additional $N^{1/3}$ scaling is caused by the Courant–Friedrich–Levy condition, because the smoothing length scales $\propto N^{-1/3}$ imposing an accordingly smaller time step. Indeed, we found that the higher resolved simulation took about 16 times longer than the low resolution run, comparable to the expected value $8.5 \cdot 6^{1/3}$. One has to keep in mind, however, that this additional $N^{1/3}$ factor only appears, if simulations with identical box sizes but different resolutions are compared, but not for simulations with identical resolution but different box sizes.

## 5.3 Efficiency of the multiple time step scheme

We finish the discussion of the performance of GRAPESPH by addressing the efficiency of the multiple time step scheme. In Fig. 1, the fraction of CPU time spent in a given subroutine is shown against the number of active particles. The figure can be roughly divided into two regions with qualitatively different behaviour. For particle numbers larger than 500 (100 for the low resolution run), the CPU time is dominated by the subroutines to calculate the density and the pressure forces. Both amount approx. 35% of the CPU time. The evaluation of gravitational forces requires 10 to 15%, other subroutines less than 10%. On recognizes, that the CPU time spent to calculate gravitational forces is about 50% larger when all particles are updated. This behaviour can be understood as follows: in general, the more diffusively distributed collisionless dark matter particles allow a larger time step than the much denser clustered gas particles. Therefore, they almost exclusively populate the highest time level. These collisionless particles, however, only consume CPU time within the gravitational force calculation. The CPU time spent for miscellaneous subroutines, increases, because only if all time levels are synchronized, subroutines are encountered which write restart files, which group gas particles in dense knots and so on.

For less than 500 (100) active particles the CPU time is clearly dominated by miscellaneous subroutines, which contribute up to 65% (42%). It turns out that nearly all time is spent to interpolate passive particles. Since the number of active particles is much less than the total number of particles, nearly every particle has to be interpolated, resulting in a constant overhead. This overhead time is, of course, larger for a larger particle number, i.e., the overhead time to propagate time levels occupied by only a few particles increases with the total number of particles. The subroutines for density and gravity also seem to have a constant overhead, whereas the CPU time spent to calculate the pressure force almost vanishes. This becomes even more clear in Fig. 2, which shows for different subroutines the mean CPU time per time step as a function of the number of active particles. As one can see, the CPU time increases almost linearly for more then 500 (100) active particles, i.e., the CPU time is proportional to the number of evaluated forces. The CPU time to update all particles slightly exceeds the linear scaling, mainly because of the additional dark matter particles (see above). For only a few active particles, the CPU time saturates at a value slightly less that 1.5 (0.7) seconds per time step. The timing of the density and gravity subroutines exhibits a similar behaviour, though the saturation appears at different particle numbers. The pressure force calculation shows an almost linear behaviour everywhere. At first glance on may be surprised, that for only a few active particles, the gravitational part requires almost 25%, although there is no active dark matter particle on such a low time level. The explanation is, that though the force has to be calculated only for a few particles, all particles have to be transmitted to GRAPE in order to calculate the gravitational force. Therefore, the time required to load GRAPE with all particles is the dominant fraction for such a small number of particles, resulting in a constant overhead. For the same reason, the density calculation requires a constant overhead, too. Since the total number of particles is twice as large as the number of gas particles, the communication time to calculate gravitational forces is roughly twice as large as that required for the density, which also is evident from Fig. 2. The pressure force calculation is almost free of overhead, since the baryonic particles are already transferred to GRAPE in order to determine the density.

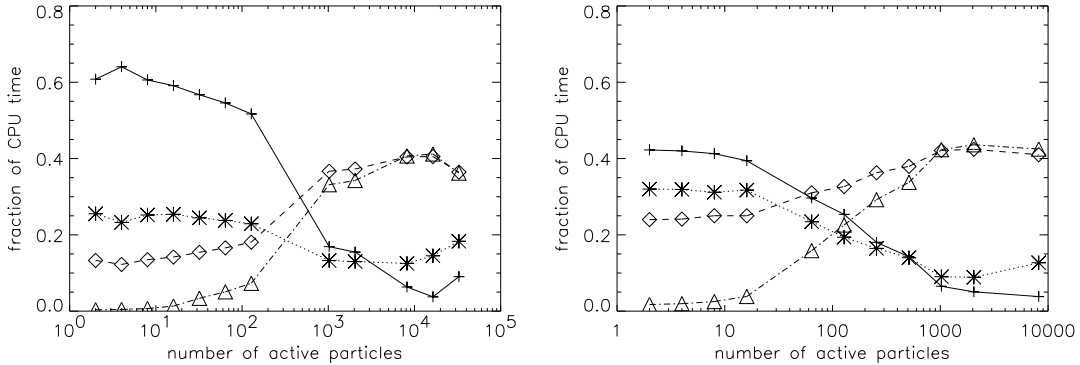

Figure 1: Fraction of CPU time spent in different subroutines as a function of the number of active baryonic particles. On the left (right) hand side, the high (low) resolution run is shown. Timings were analyzed for the redshift interval $[0.26; 0.19]$. Stars, triangles, rhombi and crosses mark the measured timings for gravitational forces, pressure forces, density and miscellaneous subroutines, respectively.

In summary, we have a nearly constant overhead of less than 1% of a complete time step in which all particles have been updated. We compare this number with TREESPH, were the overhead solely due to the tree construction can exceed 10%. From these numbers we can also estimate the gain due to the multiple time step scheme. Without such a scheme, every of the 2900 time steps on which the above analysis is based, had required $\approx 190$ CPU sec, i.e., a total CPU time of 155 CPU hours, whereas the actual computation time using the multiple time step scheme only required 12 hours, i.e., a speedup by factor of 13! Furthermore, it turns out, that more than 90% of the CPU time is spent when more than 1000 particles are active (i.e., the 3-4 highest time levels), and 30% of the CPU time on the highest time level.

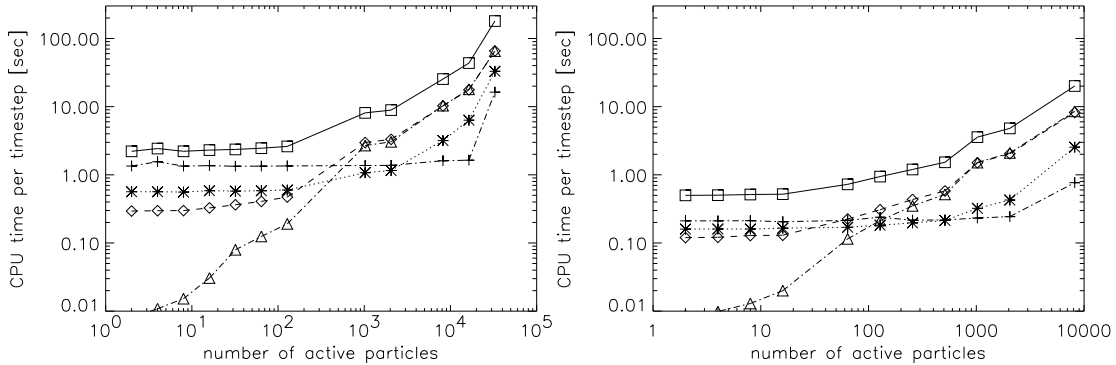

Figure 2: CPU time per time step for different subroutines as a function of the number of active particles. On the left (right) hand side, the high (low) resolution run is shown. Boxes, stars, triangles, rhombi and crosses mark the measured timings for whole program, gravitational forces, pressure forces, density and miscellaneous subroutines, respectively.

## 5.4 Comparison with other SPH codes

Comparing the performance of GRAPESPH with other SPH codes is difficult, since the required CPU time strongly depends on the simulation, e.g., whether one carries out simulations of a galaxy clusters without cooling or of a strongly cooling galaxy with star formation. In general we achieved a performance between 0.5 and 3 compared to a TREESPH code on a CRAY YMP. Compared to TREESPH the performance is best, if the gravitational part is most dominant (e.g., in simulations of forming galaxies including star formation), or if the clustering state is high. From the literature, we found four sufficiently accurate quotations of CPU timings: Katz & Gunn claim for their $2 \times 4000$ particle TREESPH-simulations of forming disk galaxies with vacuum boundaries a CPU time of 10 hours on a YMP. Similar simulations including star formation performed by Katz (1992) required $\approx 50$ hours. These timings are consistent with those estimated from our TREESPH code, which is based on a binary tree (SM93, Steinmetz & Müller 1995). Large scale structure simulations of Katz, Weinberg & Hernquist (1992) with $2 \times 32768$ particles required 200 CPU hours on a CRAY 2 to propagate the system up to a redshift of 0.4. Most recently, Frenk et al. presented P3MSPH simulations of forming galaxy clusters with $2 \times 262144$ particles. For 700 time steps, every simulation took 150 hours on a YMP.

We compare these numbers with our GRAPESPH code: The simulation of Katz & Gunn was done within 7 hours CPU time, the simulation including star formation within 20 hours, i.e., a gain in speed of a factor of 1.3–2.5. The CPU time per time step is comparable for high redshifts, when the system is only mildly non–linear. At late epochs, GRAPESPH is much more efficient the CPU time per time step being a factor of 2 (5) smaller for simulations without (with) star formation. For a simulation on larger scales ($\approx 3$ Mpc highly resolved region) where $2 \times 32000$ particles are involved, GRAPESPH took 150 hours to follow the evolution until $z = 0$, though the softening, and consequently the time step was much smaller than in the simulations of Katz, Hernquist & Weinberg (1992). However, we should note, that the GRAPESPH simulations do not assume periodic boundary conditions. To compare this simulation with that of Frenk et al., we extrapolated from our $2\times 32000$ particle simulation to a 8 times larger particle number. The timings for the density, pressure force and miscellaneous routines we assumed to increase by a factor of 10 (i.e., a scaling slightly steeper than linear as estimated above), the gravitation, however, to scale quadratically. Assuming 70 sec CPU time for the density and pressure force calculation, 35 sec for gravity, and 15 sec for miscellaneous (all timings for 32000 particles), we extrapolate a CPU time of 1.1 hours per time step or 750 hours for the complete simulation. In this case, however, the gravitational part already contributes 2/3 of the CPU time. Nevertheless, since GRAPESPH can use multiple time steps, and the clustering in the simulations of Frenk et al. (1995) is quite strong, we expect to save at least a factor of 5 ending up with a CPU time comparably to that quoted in Frenk et al. (1995).

The relative high efficiency compared to treecodes and the low one compared to P3M arises, because for treecodes the calculation of the gravitational attraction and the neighbour list requires a by a factor of a few higher fraction of the total CPU time than in case of P3M. Only for very strongly clustered system, treecodes are more efficient. The main benefit of treecodes is that it is relatively easy to implement an efficient multiple time step scheme. This is much more difficult for P3M and probably gives rise to a large overhead.

The CPU time required to perform the density and pressure force calculation is essentially the same for all kind of SPH codes differences only arising due to different ways to calculate the gravitational force and the interaction list. Since the relative weight of the SPH part in P3M is higher, the advantage of GRAPESPH is almost vanished. However, due to the high efficiency of the multiple time step scheme, GRAPESPH can still be a factor of ten faster, i.e., it can compete with P3MSPH on a supercomputer.

At the end of the discussion, we want to stress that all estimates for GRAPESPH are quite conservative. Usual SPH codes do not assume an averaged particle number of 60–70 but rather 40 or even only 32. This makes the speedup of GRAPESPH compared to P3MSPH not so impressive. Furthermore, usual turnaround times in computer centers are seldom better than 5 times the needed CPU time, whereas the turnaround times on GRAPE are close to the actual CPU time. On the other hand, the inclusion of periodic boundaries in GRAPESPH as in the simulations of Katz, Hernquist & Weinberg (1992) or Frenk et al. (1995), is still a open issue (see, however, Brieu, Summers & Ostriker 1995).

## 5.5 Future optimizations

In summary, the previous discussion showed, that GRAPESPH is to 30–40% limited by GRAPE but to 60–70% by computations of the front end. One half of the GRAPE time, however, is rather communication time than calculation time. Though the CPU time on the frontend dominates, there is no clear bottleneck. A reasonable speedup can only be achieved by tuning all parts simultaneously. Speedups of the order of 30% can be reached by the use of a faster frontend and/or a multi processor system. Because the calculation of the gravitational forces only takes a small fraction of the total time it seems not meaningful to use several boards or more sophisticated N–body algorithms adapted to GRAPE (treecodes, P3M). This statement would, however, change, if the total particle number is larger than about 200 000. In this case, however, the turnaround time just to perform the SPH–part of the simulation can exceed a few months.

Another possibility to speed up GRAPESPH is the use of a finer grained multiple time step system (White, private communication). As mentioned above, 90% of the CPU time is used when more than 1000 particles are active, (i.e., on the highest 3-4 time levels). Due to binning, on average the time steps are a factor of 1.5 smaller.

Furthermore, all CPU timings mentioned above are quite conservative. We used a code which seems to provide a good mixture between stability, accuracy, flexibility, low memory requirements and CPU time. We expect, that the code can be accelerated by several factor of 1.1 to 1.2 by using less conservative constraints. Possible gains are: (i) to use an average of 40 instead of 60 neighbours reducing the corresponding frontend end communication time accordingly. ii) Recalling the low accuracy of GRAPE, REAL*4 instead of REAL*8 arithmetic is probably more appropriate. On a SPARC 10, the frontend part can then be accelerated by about 20%. iii) Calculating the neighbour list only once and keeping it in the main memory. For an average particle number of 60 and 4 bytes per neighbour index, a 30000 gas particle simulation would require 7.2 MByte, which should not pose a severe problem on current workstations.

The situation of future developments can be seen much more optimistically, if a faster GRAPE and a faster interface is available. The announced presentation of a new generation of GRAPE with a user supplied programmable force law also seems to be attractive, especially in the context of periodic boundary conditions and GRAPEP3M. It also will provide a much larger dynamic range, comparable to REAL*4 arithmetic.

## 6. Summary and Conclusions

We present a SPH code designed to study the formation of galaxies and galaxy clusters. The code, GRAPESPH, makes intensively use of the capabilities of GRAPE, a special purpose hardware designed to efficiently solve the gravitational N–body problem. In GRAPESPH all the calculations required to determine hydrodynamical quantities and to propagate particles are done on the front end, GRAPE is only used to calculate the gravitational attraction and to determine the interaction list. In order to marry GRAPE and SPH to an efficient and accurate code, the hardware design has to be carefully analyzed. Compared to current SPH implementation, some minor modifications have to be made. We also present two further, more general modifications to standard SPH, which seem to be advantageous for cosmological applications. One modification is to steepen the gradient of the kernel, the other is to modify the artificial viscosity in order to suppress the viscosity in pure shear flows.

After a detailed and technical description how GRAPE and SPH are combined, the performance of GRAPESPH is analyzed. It turns out that the code is (almost) well balanced between calculations on the front end, on GRAPE and communication times. Because of the relative low fraction of CPU time spent to calculate gravitational interactions, a combination of GRAPESPH with more sophisticated N-body methods like treecodes or P3M seems to be not meaningful at the moment. A comparison with timings of other SPH codes running on supercomputers shows, that GRAPESPH can clearly compete with these codes. For highly clustered systems like galaxy and galaxy clusters, or for simulations involving star/galaxy formation, GRAPESPH even seems to be superior. To large part, this success is related to a multiple time step scheme, which can be applied easily and efficiently. Savings in CPU time which exceeds a factor of ten do not seem to be unreasonable. A major shortcoming of GRAPESPH is the difficulty to incorporate periodic boundary conditions.

In summary, for the price of a mid class workstation, the combination of GRAPE and SPH opens the door to applications which up to now were exclusively the domain of supercomputers. Taking into account the much shorter turnaround times on GRAPE it is even possible to attack problems which are beyond the capabilities of supercomputers. Though GRAPESPH is probably not (yet) flexible enough to attack all currently investigated problems of hydrodynamical cosmology, it is at least able to investigate a large subset.

**Acknowledgements.** It is a pleasure to thank Ewald Müller for many useful discussions about GRAPE, SPH, and hydrodynamics in general. Jun Makino is acknowledged for his advice in technical questions. This work is partially supported by the Sonderforschungsbereich *Astro-Teilchenphysik* SFB 375–95 der Deutschen Forschungsgemeinschaft.

# A. FORTRAN code skeleton to handle neighbour list

```
      SUBROUTINE density
c
      CALL sort(nsph,x(1,1),isort(1))         ! sort particles in x
      CALL g3init()                           ! acquire GRAPE board
      CALL g3setscales(xmin,cmmin)            ! set scaling
      CALL g3seteps2(xmin*xmin)               ! no softening
c
      CALL g3setn(nsph)                       ! pass positions and masses
      DO j = 1, nsph                          ! to the board
        CALL g3setxj(i-1, x(1,i) )
        CALL g3setmj(i-1, m (i) )
      ENDDO
c
      DO j8 = 1, nsph, nchip                  ! loop over particles
c
        nsub = nsph
        DO i = 1, nchip                       ! initialize neighbour
c                                             ! list to zero
          j= isort(i+j8-1)
          neigh (j) = 0
          istart(i) = 1                       ! neighbour list offset
c
          h2 = (1.3*h(j))**2                  ! search radius
          CALL g3seth2(h2,i-1)
c
          y(1,i) = x(1,j)                     ! position, for which
          y(2,i) = x(2,j)                     ! neighbours should be
          y(3,i) = x(3,j)                     ! found
        ENDDO
c
 42     CONTINUE                              ! start iteration
c
        IF (nsub.EQ.nsph)
          CALL g3frc(y,acc,pot,nchip)         ! start GRAPE
        ELSE
          CALL g3setn(nsub+1)                 ! set chip1 to the nsub+1
          CALL g3setxj(nsub,x(1,j8))          ! particle to guarantee at
          CALL g3setmj(nsub,m (j8) )          ! least one neighbour
c
          CALL g3frc(y,acc,pot,nchip)         ! start GRAPE
c
          CALL g3setxj(nsub,x(1,nsub+1))      ! restore original particle
          CALL g3setmj(nsub,m (nsub+1))       ! nsub+1
        ENDIF
```

```fortran
      ntot = g3readnbl(0)                    ! get number of
c                                            ! disjunct neighbours
      IF (ntot.LT.1023) THEN                 ! handle overflow
c                                            ! was successful
        DO i = 1, nchip
          j = isort(i+j8-1)
          nt = g3getnbl(i-1, index(k(i),i))  ! read neighbour list
c
          if ((index(k(i)-1+nt,i).GE.nsub)   ! omit particle nsub+1
     &          .AND.(nt.GE.1)) nt = nt - 1  ! from neighbour list
c
          neigh(j) = neigh(j) + nt
          CALL densty(j, neigh(j), k(1,i))   ! get density, force etc.
        ENDDO
c
      ELSE                                   ! neighbour search
c
        nsubnew = nsub
        DO i = 1, nchip
          j = isort(i+j8-1)
          nt = g3getnbl(i-1, k(istart(i),i)) ! read neighbour list
c
          IF ((k(istart(i)-1+nt,i).GE.nsub)  ! omit particle nsub+1
     &          .AND.(nt.GE.1)) nt = nt - 1  ! from neighbour list
c
          neigh(j) = neigh(j) + nt
          IF (nt.GT.0) nsubnew =
     &       MIN(nsubnew,k(istart(i),i) )
          istart(i) = istart(i) + nt         ! set new offset
        ENDDO
c
        nsub = nsubnew                       ! define subset for
        IF (nsub.GT.0) GOTO 42               ! the next iteration
c
        DO i = 1, nchip
          j = isort(i+j8-1)
          CALL densty(j, neigh(j), k(1,i))   ! get density, force etc.
        ENDDO
c
      ENDIF
c
      ENDDO
c
      CALL g3free                            ! release GRAPE board
c
      RETURN
      END
```